\documentclass[a4paper,UKenglish,numberwithinsect]{lipics-v2018}

\usepackage{amsmath,amssymb,graphicx,pict2e}
\usepackage{microtype}%if unwanted, comment out or use option "draft"
\title{Dispersing obnoxious facilities on a graph}
\titlerunning{Dispersing obnoxious facilities on a graph}

\author{Alexander Grigoriev}{Department of Quantitative Economics, Maastricht University, The Netherlands}
  {a.grigoriev@maastrichtuniversity.nl}{}{}
\author{Tim A. Hartmann}{Department of Computer Science, RWTH Aachen, Germany}{hartmann@algo.rwth-aachen.de}{}{}
\author{Stefan Lendl}{Institute of Discrete Mathematics, TU Graz, Austria}{lendl@math.tugraz.at}{}
  {Supported by the Austrian Science Fund (FWF): W1230, Doctoral Program ``Discrete Mathematics''.}
\author{Gerhard J. Woeginger}{Department of Computer Science, RWTH Aachen, Germany}{woeginger@algo.rwth-aachen.de}{}
  {Supported by the DFG RTG 2236 ``UnRAVeL''.}
\authorrunning{A. Grigoriev, T.A. Hartmann, S. Lendl, and G.J. Woeginger}

\Copyright{Alexander Grigoriev, Tim A. Hartmann, Stefan Lendl, and Gerhard J. Woeginger}
\subjclass{Mathematics of computing $\rightarrow$ Graph theory, 
Theory of computation $\rightarrow$ Graph algorithms analysis,
Theory of computation $\rightarrow$ Discrete optimization}
\acknowledgements{The research in this paper has been started during the Lorentz Center Workshop 
on Fixed-Parameter Computational Geometry, which was organized in May 2018 in Leiden.
We thank the Lorentz Center for its hospitality.
Alexander Grigoriev and Gerhard Woeginger acknowledge helpful discussions with Fedor Fomin, Petr Golovach and Jesper Nederlof.}

\EventEditors{John Q. Open and Joan R. Access}
\EventNoEds{2}
\EventLongTitle{42nd Conference on Very Important Topics (CVIT 2019)}
\EventShortTitle{CVIT 2019}
\EventAcronym{CVIT}
\EventYear{2019}
\EventDate{December 24--27, 2019}
\EventLocation{Little Whinging, United Kingdom}
\EventLogo{}
\SeriesVolume{42}
\ArticleNo{XX}
\nolinenumbers %uncomment to disable line numbering
\hideLIPIcs  %uncomment to remove references to LIPIcs series (logo, DOI, ...) 
\newcommand{\disp}[2]{#1\mbox{\rm-Disp}(#2)}
\newcommand{\indset}{{\sc Cubic-Ind-Set}}

\newcommand{\xodd}{X_{\ge3}}
\newcommand{\vic}{\text{vic}}

\begin{document}
\maketitle

\begin{abstract}
We study a continuous facility location problem on a graph where all edges have unit
length and where the facilities may also be positioned in the interior of the edges.
The goal is to position as many facilities as possible subject to the condition
that any two facilities have at least distance $\delta$ from each other.

We investigate the complexity of this problem in terms of the rational parameter $\delta$.
The problem is polynomially solvable, if the numerator of $\delta$ is $1$ or $2$,
while all other cases turn out to be NP-hard.
\end{abstract}

\newpage
%%%%%%%%%%%%%%%%%%%%%%%%%%%%%%%%%%%%%%%%%%%%%%%%%%%%%%%%%%%%%%%%%%%%%%%%%
%%%%%%%%%%%%%%%%%%%%%%%%%%%%%%%%%%%%%%%%%%%%%%%%%%%%%%%%%%%%%%%%%%%%%%%%%
%%%%%%%%%%%%%%%%%%%%%%%%%%%%%%%%%%%%%%%%%%%%%%%%%%%%%%%%%%%%%%%%%%%%%%%%%
\section{Introduction}
%%%%%%%%%%%%%%%%%%%%%%%%%%%%%%%%%%%%%%%%%%%%%%%%%%%%%%%%%%%%%%%%%%%%%%%%%
A large part of the facility location literature deals with \emph{desirable} facilities that people
like to have nearby, such as service centers, police departments, fire stations, and warehouses. 
However, there also do exist facilities that are \emph{undesirable} and \emph{obnoxious}, such as nuclear 
reactors, garbage dumps, chemical plants, military installations, and high security penal institutions.
A standard goal in location theory is to spread out such obnoxious facilities and to avoid their 
accumulation and concentration in a small region; see for instance Erkut \& Neuman \cite{ErkNeu1989} 
and Cappanera \cite{Cappanera1999} for comprehensive surveys on this topic.

In this paper, we investigate the location of obnoxious facilities in a metric space whose topology
is determined by a graph.
Formally, let $G=(V,E)$ be an undirected connected graph, where every edge is rectifiable and has unit length.
Let $P(G)$ denote the continuum set of points on all the edges in $E$ together with all the vertices in $V$.
For two points $p,q\in P(G)$, we denote by $d(p,q)$ the length of a shortest path connecting $p$ and $q$ 
in the graph.
A subset $S\subset P(G)$ is said to be \emph{$\delta$-dispersed} for some positive real number $\delta$,
if any two points $p,q\in S$ with $p\ne q$ are at distance $d(p,q)\ge\delta$ from each other.
Our goal is to compute for a given graph $G=(V,E)$ and a given positive real number $\delta$ a maximum 
cardinality subset $S\subset P(G)$ that is $\delta$-dispersed. 
Such a set $S$ is called an \emph{optimal} $\delta$-dispersed set, and $|S|$ is called the 
\emph{$\delta$-dispersion number} $\disp{\delta}{G}$ of the graph $G$.

\subsection*{Known and related results.}
Obnoxious facility location goes back to the seminal articles of Goldman \& Dearing \cite{GolDea1975} 
from 1975 and Church \& Garfinkel \cite{ChuGar1978} from 1978.
The area actually covers a wide variety of problem variants and models; some models specify a geometric 
setting, while other models use a graph-theoretic setting.

For example, Abravaya \& Segal \cite{AbrSeg2010} consider a purely geometric variant of obnoxious facility 
location, where a maximum cardinality set of obnoxious facilities has to be placed in a rectangular region, such 
that their pairwise distance as well as the distance to a fixed set of demand sites is above a given threshold.
As another example we mention the graph-theoretic model of Tamir \cite{Tamir1991}, where every edge $e\in E$ 
of the underlying graph $G=(V,E)$ is rectifiable and has a given edge-dependent length $\ell(e)$.
Tamir discusses the complexity and approximability of various optimization problems with various objective functions.
One consequence of \cite{Tamir1991} is that if the graph $G$ is a tree, then the value $\disp{\delta}{G}$ 
can be computed in polynomial time.
Segal \cite{Segal2003} locates a single obnoxious facility on a network under various objective functions,
such as maximizing the smallest distance from the facility to the clients on the network or maximizing the
total sum of the distances between facility and clients. 

Megiddo \& Tamir \cite{MegTam1983} consider the covering problem that is dual to the $\delta$-dispersion 
packing problem: Given a graph $G=(V,E)$ with rectifiable unit-length edges, find a minimum cardinality 
subset $S\subset P(G)$ such that every point in $P(G)$ is at distance at most $\delta$ from one of the 
facilities in $S$.
Among many other results \cite{MegTam1983} shows that this covering problem is NP-hard for $\delta=2$.  

Finally, we mention the work of 
Gawrychowski, Krasnopolsky, Mozes \& Weimann \cite{GawKraMozWei2017} who study the problem variant where
the points in the dispersed set $S$ must be vertices of the graph $G$.
They show that for a given tree $G$ and a given integer $k$, one can compute in linear time the largest
possible value $\delta$ for which there exists a $\delta$-dispersed set $S$ of size $|S|=k$.

\subsection*{Our results.}
We provide a complete picture of the complexity of computing the $\delta$-dispersion number for
connected graphs $G=(V,E)$ and positive rational numbers $\delta$.
\begin{itemize}
\item If $\delta=1/b$ for some integer $b$, then the $\delta$-dispersion number of $G$ can be 
written down without really looking at the structure of the graph: 
If $G$ is a tree then $\disp{\delta}{G}=b|E|+1$, and if $G$ is not a tree then $\disp{\delta}{G}=b|E|$.
\item If $\delta=2/b$ for some integer $b$, then $\disp{\delta}{G}$ can be computed in polynomial time.
The algorithm uses the Edmonds-Gallai decomposition of $G$ and reformulates the problem as a submodular 
optimization problem.
\item If $\delta=a/b$ for integers $a$ and $b$ with $a\ge3$ and $\gcd(a,b)=1$, then the computation
of $\disp{\delta}{G}$ is an NP-hard problem.
\end{itemize}
The rest of the paper is organized as follows.
Section~\ref{sec:prel} summarizes the basic notations and states several technical observations. 
Section~\ref{sec:np} presents the NP-hardness results. 
The reductions are essentially based on routine methods, but need to resolve certain number-theoretic issues.
Our technical main contribution is the polynomial time algorithm for the case $\delta=2$ as developed 
in Section~\ref{sec:delta=2}; this result is heavily based on tools from matching theory.
Section~\ref{sec:polynomial} summarizes the polynomially solvable special cases and provides 
additional structural insights.

%%%%%%%%%%%%%%%%%%%%%%%%%%%%%%%%%%%%%%%%%%%%%%%%%%%%%%%%%%%%%%%%%%%%%%%%%
%%%%%%%%%%%%%%%%%%%%%%%%%%%%%%%%%%%%%%%%%%%%%%%%%%%%%%%%%%%%%%%%%%%%%%%%%
%%%%%%%%%%%%%%%%%%%%%%%%%%%%%%%%%%%%%%%%%%%%%%%%%%%%%%%%%%%%%%%%%%%%%%%%%
\section{Notation and technical preliminaries}
\label{sec:prel}
%%%%%%%%%%%%%%%%%%%%%%%%%%%%%%%%%%%%%%%%%%%%%%%%%%%%%%%%%%%%%%%%%%%%%%%%%
All graphs in this paper are undirected and connected, and all edges have unit length.
Throughout the paper we use the word \emph{vertex} in the graph-theoretic sense, and we use
the word \emph{point} to denote the elements of the geometric structure $P(G)$.
For a graph $G=(V,E)$ and a subset $V'\subseteq V$, we denote by $G[V']$ the subgraph induced by $V'$.
For an integer $c\ge1$, the \emph{$c$-subdivision} of $G$ is the graph that results from $G$ 
by subdividing every edge in $E$ by $c-1$ new vertices into $c$ new edges.

For an edge $e=\{u,v\}$ and a real number $\lambda$ with $0\le\lambda\le1$, we denote by $p(u,v,\lambda)$
the point on $e$ that has distance $\lambda$ from vertex $u$.
Note that $p(u,v,0)=u$ and $p(u,v,1)=v$, and note that point $p(u,v,\lambda)$ coincides with point
$p(v,u,1-\lambda)$; hence we will sometimes assume without loss of generality that $\lambda\le1/2$.

%%%%%%%%%%%%%%%%%
\begin{lemma}
\label{le:scaling}
Let $G$ be a graph, let $c\ge1$ be an integer, and let $G'$ be the $c$-subdivision of $G$.
Then for every $\delta>0$, the $\delta$-dispersed sets in $G$ are in one-to-one correspondence 
with the $(c\cdot\delta)$-dispersed sets in $G'$.
In particular, $\disp{\delta}{G}=\disp{(c\cdot\delta)}{G'}$.
\end{lemma}
%%%%%%%%%%%%%%%%%
\proof
Every point $p(u,v,\lambda)$ in $P(G)$ translates into a corresponding point in $P(G')$ that lies 
on the subdivided edge between $u$ and $v$ and is at distance $c\cdot\lambda$ from vertex $u$.
\qed

\medskip
Lemma~\ref{le:scaling} has many useful consequences, as for instance the following:
%%%%%%%%%%%%%%%%%
\begin{lemma}
\label{le:monotonicity}
Let $\delta>0$ and let $c\ge1$ be an integer.
\begin{itemize}
\item If the problem of computing the $\delta$-dispersion number is NP-hard,
then also the problem of computing the $(c\cdot\delta)$-dispersion number is NP-hard.
\item If the problem of computing the $(c\cdot\delta)$-dispersion number is polynomially solvable,
then also the problem of computing the $\delta$-dispersion number is polynomially solvable.
\end{itemize}
\end{lemma}
%%%%%%%%%%%%%%%%%
\proof
By Lemma~\ref{le:scaling} the $c$-subdivision of a graph yields a polynomial time reduction from 
computing $\delta$-dispersions to computing $(c\cdot\delta)$-dispersions.
\qed

\medskip
For integers $\ell$ and $k$, the rational number $\ell/k$ is called \emph{$k$-simple}.
A set $S\subseteq P(G)$ is $k$-simple, if for every point $p(u,v,\lambda)$ in $S$ the number
$\lambda$ is $k$-simple.
%%%%%%%%%%%%%%%%%
\begin{lemma}
\label{le:halfintegral}
Let $\delta=a/b$ with integers $a$ and $b$, and let $G=(V,E)$ be a graph.
Then there exists an optimal $\delta$-dispersed set $S^*$ that is $2b$-simple.
\end{lemma}
%%%%%%%%%%%%%%%%%
\proof
We first handle the cases with $b=1$, so that $\delta$ is integer. 
Consider an optimal $\delta$-dispersed set $S$ for graph $G$.
Note that for every vertex $u$, at most one point $p(u,v,\lambda)$ with $v\in V$ and $0\le\lambda<1/2$
is in $S$. 
For every point $p=p(u,v,\lambda)$ with $0\le\lambda\le1/2$ in $S$, we put a corresponding 
point $p^*$ into set $S^*$:
If $0\le\lambda<1/2$ then $p^*=p(u,v,0)$, and if $\lambda=1/2$ then $p^*=p(u,v,1/2)$.
As all points in the resulting set $S^*$ are either vertices or midpoints of edges,
we get that $S^*$ is $2$-simple.
We claim that $S^*$ is still $\delta$-dispersed:
Consider two distinct points $p^*$ and $q^*$ in $S^*$.
Note that $d(p,p^*)<1/2$ and $d(q,q^*)<1/2$ by construction.
\begin{itemize}
\item If $p^*$ and $q^*$ both are vertices in $V$, then the distance $d(p^*,q^*)$ is integer.
By the triangle inequality $d(p,q)\le d(p,p^*)+d(p^*,q^*)+d(q^*,q)$.
As the left hand side in this inequality is at least the integer $\delta$ and as its right hand side 
is strictly smaller than the integer $d(p^*,q^*)+1$, we conclude $d(p^*,q^*)\ge\delta$.
\item If $p^*$ and $q^*$ both are midpoints of edges, then $p=p^*$ and $q=q^*$ yields $d(p^*,q^*)\ge\delta$.
\item If $p^*$ is a vertex and $q^*$ is the midpoint of some edge, then $d(p^*,q^*)=D+1/2$ for
some integer $D$. 
The triangle inequality together with $p=p^*$ yields $\delta\le d(p,q)=d(p^*,q)\le d(p^*,q^*)+d(q^*,q)<D+1$.
This implies $D\ge\delta$, so that $d(p^*,q^*)\ge\delta+1/2$.  
\end{itemize}
Since $S$ and $S^*$ have the same cardinality, we conclude that $S^*$ is an optimal $\delta$-dispersed 
set that is $2$-simple, exactly as desired.

In the cases where $\delta=a/b$ for some integer $b\ge2$, we consider the $b$-subdivision $G'$ of $G$.
By the above discussion, $G'$ possesses an optimal $a$-dispersed set $S'$ that is $2$-simple.
Then Lemma~\ref{le:scaling} translates $S'$ into an optimal $\delta$-dispersed 
set $S$ for $G$ that is $2b$-simple.
\qed

%%%%%%%%%%%%%%%%%%%%%%%%%%%%%%%%%%%%%%%%%%%%%%%%%%%%%%%%%%%%%%%%%%%%%%%%%
%%%%%%%%%%%%%%%%%%%%%%%%%%%%%%%%%%%%%%%%%%%%%%%%%%%%%%%%%%%%%%%%%%%%%%%%%
%%%%%%%%%%%%%%%%%%%%%%%%%%%%%%%%%%%%%%%%%%%%%%%%%%%%%%%%%%%%%%%%%%%%%%%%%
\section{NP-completeness results}
\label{sec:np}
%%%%%%%%%%%%%%%%%%%%%%%%%%%%%%%%%%%%%%%%%%%%%%%%%%%%%%%%%%%%%%%%%%%%%%%%%
In this section we present our NP-hardness proofs for computing the $\delta$-dispersion number. 
All proofs are done through polynomial time reductions from the following NP-hard
variant of the independent set problem; see Garey \& Johnson \cite{GarJoh1979}.
%%%%%%%%%%%%%%%%%
\begin{quote}
Problem: Independent Set in Cubic Graphs ({\indset})
\\[1.0ex]
Instance: An undirected, connected graph $H=(V_H,E_H)$ in which every vertex is adjacent to
exactly three other vertices; an integer bound $k$.
\\[1.0ex]
Question:
Does $H$ contain an independent set $I$ with $|I|\ge k$ vertices?
\end{quote}
%%%%%%%%%%%%%%%%%
Throughout this section we consider a fixed rational number $\delta=a/b$, where $a$ and $b$ are positive
integers that satisfy $\gcd(a,b)=1$ and $a\ge3$.
Section~\ref{ssec:np.1} the cases with odd  numerators $a\ge3$, and
Section~\ref{ssec:np.2} the cases with even numerators $a\ge4$.
It is instructive to verify that our arguments do not work for the cases with $a=1$ and $a=2$,
as our gadgets and our arguments break down at various places. 

%%%%%%%%%%%%%%%%%%%%%%%%%%%%%%%%%%%%%%%%%%%%%%%%%%%%%%%%%%%%%%%%%%%%%%%%%
\subsection{NP-hard cases with odd numerator}
\label{ssec:np.1}
%%%%%%%%%%%%%%%%%%%%%%%%%%%%%%%%%%%%%%%%%%%%%%%%%%%%%%%%%%%%%%%%%%%%%%%%%
Throughout this section we consider a fixed rational number $\delta=a/b$ where $\gcd(a,b)=1$ and where
$a\ge3$ is an odd integer.
For the NP-hardness proof, we first determine four positive integers $x_1,y_1,x_2,y_2$ 
that satisfy the following equations \eqref{eq:Bezout.1} and \eqref{eq:Bezout.2}.
%%%%%%%%%%%%%%%%%
\begin{eqnarray}
2b\cdot x_1 - 2a\cdot y_1 &=& a-1 \label{eq:Bezout.1} \\[0.5ex]
 b\cdot x_2 -  a\cdot y_2 &=&   1 \label{eq:Bezout.2}
\end{eqnarray}
%%%%%%%%%%%%%%%%%
Note that the value $a-1$ on the right hand side of equation \eqref{eq:Bezout.1} is even, and 
hence is divisible by the greatest common divisor $\gcd(2b,2a)=2$ of the coefficients in the left hand side.
With this, B\'ezout's lemma yields the existence of positive integers $x_1$ and $y_1$ that satisfy \eqref{eq:Bezout.1}.  
B\'ezout's lemma also yields the existence of positive integers $x_2$ and $y_2$ in 
equation \eqref{eq:Bezout.2}, as the coefficients in the left hand are relatively prime.

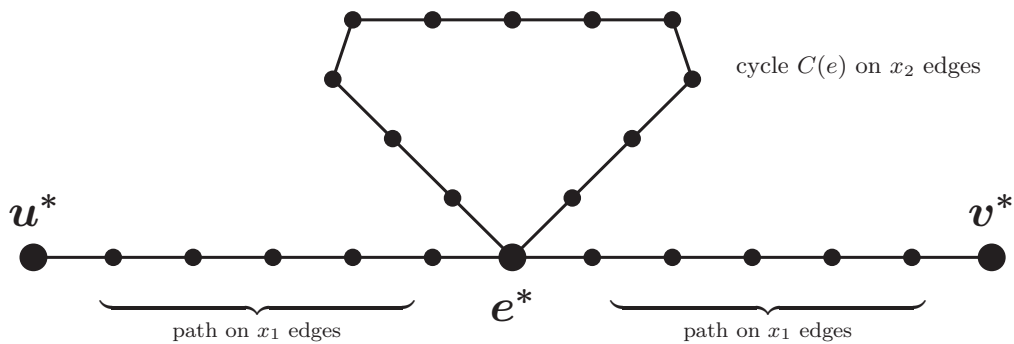
\begin{figure}[tb]
\begin{center}
\unitlength=1.05mm
\begin{picture}(120,50)(0,0)
\linethickness{0.40mm}
\put(  0,10){\line(1,0){120}}
\multiput(10,10)(10,0){11}{\circle*{2.2}}
\put(  0,10){\circle*{3.5}}
\put( 60,10){\circle*{3.5}}
\put(120,10){\circle*{3.5}}
\put(  0,16){\makebox(0,0)[cc]{\LARGE\boldmath $u^*$}}
\put(120,16){\makebox(0,0)[cc]{\LARGE\boldmath $v^*$}}
\put( 60, 4){\makebox(0,0)[cc]{\LARGE\boldmath $e^*$}}
\put( 28, 2){\makebox(0,0)[cc]{\Large$\underbrace{\hspace{10em}}_{\text{path on $x_1$ edges}}$}}
\put( 92, 2){\makebox(0,0)[cc]{\Large$\underbrace{\hspace{10em}}_{\text{path on $x_1$ edges}}$}}
\put( 60,10){\line( 1,1){23}}
\put( 60,10){\line(-1,1){23}}
\put( 40,40){\line( 1,0){40}}
\put( 40,40){\line(-1,-3){2.5}}
\put( 80,40){\line( 1,-3){2.5}}
\multiput(60,10)( 7.5,7.5){4}{\circle*{2.2}}
\multiput(60,10)(-7.5,7.5){4}{\circle*{2.2}}
\multiput(40,40)(  10,  0){5}{\circle*{2.2}}
\put( 88,34){\makebox(0,0)[lc]{\small cycle $C(e)$ on $x_2$ edges}}
\end{picture}
\end{center}
\caption{The edge $e=\{u,v\}$ in the instance of {\indset} translates into three vertices 
$u^*$, $e^*$, $v^*$ in the dispersion instance, together with two paths and one cycle.}
\label{fig:gadget}
\bigskip
\end{figure}

Our reduction now starts from an arbitrary instance $H=(V_H,E_H)$ and $k$ of {\indset}, 
and constructs a corresponding dispersion instance $G=(V_G,E_G)$ from it.
\begin{itemize}
\item
For every vertex $v\in V_H$, we create a corresponding vertex $v^*$ in $V_G$.
\item
For every edge $e=\{u,v\}\in E_H$, we create a corresponding vertex $e^*$ in $V_G$.
\item
For every edge $e=\{u,v\}\in E_H$, we create 
(i) a path with $x_1$ edges that connects vertex $u^*$ to vertex $e^*$,
(ii) another path with $x_1$ edges that connects $v^*$ to $e^*$,
and (iii) a cycle $C(e)$ with $x_2$ edges that runs through vertex $e^*$.
\end{itemize}
This completes the description of the graph $G=(V_G,E_G)$;
see Figure~\ref{fig:gadget} for an illustration.
We claim that graph $H$ contains an independent set of size $k$, if and only if
$\disp{(a/b)}{G}\ge k+(2y_1+y_2)|E_H|$.

%%%%%%%%%%%%%%%%%%%%%%%%%%%%%%%%%%%%
%%%%%%%%%%%%%%%%%%%%%%%%%%%%%%%%%%%%

%%%%%%%%%%%%%%%%%
\begin{lemma}
\label{le:hard.1}
If graph $H$ contains an independent set of size $k$, 
then the $(a/b)$-dispersion number of graph $G$ is at least $k+(2y_1+y_2)|E_H|$.
\end{lemma}
%%%%%%%%%%%%%%%%%
\proof
Let $I$ be an independent set of size $k$ in graph $H=(V_H,E_H)$.
We construct from $I$ a $\delta$-dispersed set $S\subset P(G)$ as follows.
Let $u\in V_H$ be a vertex, and let $e_1,e_2,e_3$ be the three edges in $E_H$ that are incident to $u$.
\begin{itemize}
\item
If $u\in I$, then we put point $u^*$ into $S$.
On each of the three paths that connect vertex $u^*$ respectively to vertex $e_i^*$ ($i=1,2,3$), 
we select $y_1$ further points for $S$.
The first selected point is at distance $\delta$ from $u^*$, and every further selected point is 
at distance $\delta=a/b$ from the preceding selected point.
By equation \eqref{eq:Bezout.1}, on each of the three paths the distance from the final selected point 
to point $e_i^*$ ($i=1,2,3$) then equals $(a-1)/(2b)$.
\item
If $u\notin I$, then on each of the three paths between $u^*$ and $e_i^*$ ($i=1,2,3$) we select $y_1$ points for $S$.
The first selected point is at distance $\delta/2=a/(2b)$ from $u^*$, and every further selected point is 
at distance $\delta$ from the preceding selected point.
By equation \eqref{eq:Bezout.1}, the distance from the final selected point to point $e^*$ then equals $(2a-1)/(2b)$.
\end{itemize}
Furthermore, for every edge $e\in E_H$ we select $y_2$ points from the cycle $C(e)$ for $S$: 
\begin{itemize}
\item
We start in point $e^*$ and traverse $C(e)$ in clockwise direction.
The first selected point is at distance $(a+1)/(2b)$ from point $e^*$, and every further
selected point is at distance $\delta$ from the preceding selected point.
By equation \eqref{eq:Bezout.2}, the distance from the final selected point to point $e^*$ then equals $(a+1)/(2b)$.
\end{itemize}
This completes the construction of set $S$.
Now let us count the points in $S$.
First, there are the $k$ points $u^*\in S$ for which $u\in I$.
Furthermore, for every edge $e=\{u,v\}\in E_H$ there are $2y_1$ points in $S$ that lie on the two paths 
from $u^*$ to $e^*$ and from $e^*$ to $v^*$.
Finally, for every edge $e\in E_H$ there are $y_2$ points that lie on the cycle $C(e)$.
Altogether, this yields the desired size $k+(2y_1+y_2)|E_H|$ for $S$.

It remains to verify that the point set $S$ is $\delta$-dispersed.
By construction, the points selected from each path are at distance at least $\delta$ from each other,
and the same holds for the points selected from each cycle.
If vertex $u^*$ is in $S$, then all selected points on the three incident paths are at distance at least $\delta$ from $u^*$.
If vertex $u^*$ is not in $S$, then the first selected point on every path is at distance $\delta/2$ from $u^*$,
so that these points are pairwise at distance at least $\delta$ from each other.
Hence the only potential trouble could arise in the neighborhood of point $e^*$, where paths and cycles are glued together.
Every selected point on $C(e)$ is at distance at least $(a+1)/(2b)$ from point $e^*$.
Every selected point on some path from $u^*$ to $e^*$ is at distance at least $(a-1)/(2b)$ from $e^*$ if $u\in I$
and is at distance at least  $(2a-1)/(2b)$ if $u\notin I$.
Since for any edge $e=\{u,v\}\in E_H$ at most one of the end vertices $u$ and $v$ is in $I$, 
at most one selected point can be at distance $(a-1)/(2b)$ from $e^*$, and all other points are at distance at least
$(a+1)/(2b)$ from $e^*$.
Hence $S$ is indeed $\delta$-dispersed.
\qed

%%%%%%%%%%%%%%%%%
\begin{lemma}
\label{le:hard.2}
If the $(a/b)$-dispersion number of graph $G$ is at least $k+(2y_1+y_2)|E_H|$,
then graph $H$ contains an independent set of size $k$.
\end{lemma}
%%%%%%%%%%%%%%%%%
\proof
Let $S$ be an $(a/b)$-dispersed set of size $k+(2y_1+y_2)|E_H|$.
By Lemma~\ref{le:halfintegral} we assume that for every point $p(u,v,\lambda)$ in $S$, 
the denominator of the rational number $\lambda$ is $2b$.

For an edge $e=\{u,v\}\in E_H$, let us consider its corresponding path $\pi$ on $x_1$ edges that 
connects vertex $u^*$ to vertex $e^*$.  
Suppose that there is some point $p$ in $S\cap\pi$ with $d(p,e^*)\le(a-2)/(2b)$.
Then by Equation \eqref{eq:Bezout.2}, set $S$ will contain at most $y_2-1$ points from the cycle $C(e)$.
In this case we restructure $S$ as follows:
We remove point $p$ together with the at most $y_2-1$ points on cycle $C(e)$ from $S$, and instead 
insert $y_2$ points into $S$ that are $\delta$-dispersed on $C(e)$ and that all are at distance at 
least $(a+1)/(2b)$ from $e^*$.
As this restructuring does not decrease the size of $S$, we will from now on assume without
loss of generality that $d(p,e^*)\ge(a-1)/(2b)$ holds for every point $p\in S\cap\pi$.

Now let us take a closer look at the points in $S\cap\pi$.
Equation \eqref{eq:Bezout.1} can be rewritten into $x_1=y_1\delta+(a-1)/(2b)$, which 
yields $|S\cap\pi|\le y_1+1$.
\begin{itemize}
\item In the equality case $|S\cap\pi|=y_1+1$, we must have $u^*\in S$ and also the point 
on $\pi$ at distance $(a-1)/(2b)$ from $e^*$ must be in $S$.
\item In case $|S\cap\pi|\le y_1$, there is ample space for picking $y_1$ points from $\pi$ that 
are $\delta$-dispersed and that are at distance at least $\delta/2$ from $u^*$ and at distance 
at least $\delta/2$ from $e^*$.
Hence we will from now on assume $|S\cap\pi|=y_1$ in these cases.
\end{itemize}

Now let us count:
Set $S$ contains exactly $y_1$ interior points from every path $\pi$, and altogether there are $2|E_H|$ such paths.
Set $S$ contains exactly $y_2$ points from every cycle $C(e)$, and altogether there are $|E_H|$ such cycles.
Since $|S|\ge k+(2y_1+y_2)|E_H|$, this means that $S$ must contain at least $k$ further points on
vertices $u^*$ with $u\in V_H$.
The corresponding subset of $V_H$ is called $I$. 

Finally, we claim that this set $I$ with $|I|\ge k$ forms an independent set in graph $H$.
Suppose for the sake of contradiction that there is an edge $e=\{u,v\}\in E_H$ with $u^*\in I$
and $v^*\in I$.
Consider the two paths that connect $u^*$ to $e^*$ and $v^*$ to $E^*$.
By the above discussion, $S$ then contains two points at distance $(a-1)/(2b)$ from $e^*$.
As these two points are then at distance at most $(a-1)/b<\delta$ from each other,
we arrive at the desired contradiction.
\qed

\bigskip
The statements in Lemma~\ref{le:hard.1} and in~\ref{le:hard.2} yield the following theorem.
%%%%%%%%%%%%%%%%%
\begin{theorem}
\label{th:npc-odd}
Let $a$ and $b$ be positive integers with $\gcd(a,b)=1$ and odd $a\ge3$.
Then it is NP-hard to compute the $(a/b)$-dispersion number of a graph $G$.
\end{theorem}
%%%%%%%%%%%%%%%%%

%%%%%%%%%%%%%%%%%%%%%%%%%%%%%%%%%%%%%%%%%%%%%%%%%%%%%%%%%%%%%%%%%%%%%%%%%
\subsection{NP-hard cases with even numerator}
\label{ssec:np.2}
%%%%%%%%%%%%%%%%%%%%%%%%%%%%%%%%%%%%%%%%%%%%%%%%%%%%%%%%%%%%%%%%%%%%%%%%%
In this section we consider a fixed rational number $\delta=a/b$ where $\gcd(a,b)=1$ and where
$a\ge4$ is an even integer.
The NP-hardness argument is essentially a minor variation of the argument in 
Section~\ref{ssec:np.1} for the cases with odd numerators.
Therefore, we will only explain the modifications, and leave all further details to the reader.

The NP-hardness proof in Section~\ref{ssec:np.1} is centered around the four positive 
integers $x_1,y_1,x_2,y_2$ introduced in equations \eqref{eq:Bezout.1} and \eqref{eq:Bezout.2}.
We perform the same reduction from {\indset} as in Section~\ref{ssec:np.1} but with positive 
integers $x_1,y_1,x_2,y_2$ that satisfy the following equations \eqref{eq:Bezout.1b} and \eqref{eq:Bezout.2b}.
%%%%%%%%%%%%%%%%%
\begin{eqnarray}
2b\cdot x_1 - 2a\cdot y_1 &=& a-2  \label{eq:Bezout.1b} \\[0.5ex]
 b\cdot x_2 -  a\cdot y_2 &=&   2  \label{eq:Bezout.2b}
\end{eqnarray} 
%%%%%%%%%%%%%%%%%
In \eqref{eq:Bezout.1b}, the right hand side $a-2$ is even and divisible by the greatest
common divisor of the coefficients in the left hand side.
In \eqref{eq:Bezout.2b}, the coefficients in the left hand are relatively prime.
Therefore B\'ezout's lemma can be applied to both equations.

The graph $G=(V_G,E_G)$ is defined as before, with a vertex $v^*$ for every $v\in V_H$
and a vertex $e^*$ for every $e\in E_H$, with paths on $x_1$ edges and cycles $C(e)$
on $x_2$ edges.
The arguments in Lemma~\ref{le:hard.1} and~\ref{le:hard.2} can easily be adapted 
and yield the following theorem.
%%%%%%%%%%%%%%%%%
\begin{theorem}
\label{th:npc-even}
Let $a$ and $b$ be positive integers with $\gcd(a,b)=1$ and even $a\ge4$.
Then it is NP-hard to compute the $(a/b)$-dispersion number of a graph $G$.
\end{theorem}
%%%%%%%%%%%%%%%%%

%%%%%%%%%%%%%%%%%%%%%%%%%%%%%%%%%%%%%%%%%%%%%%%%%%%%%%%%%%%%%%%%%%%%%%%%%
\subsection{Containment in NP}
\label{ssec:np.4}
%%%%%%%%%%%%%%%%%%%%%%%%%%%%%%%%%%%%%%%%%%%%%%%%%%%%%%%%%%%%%%%%%%%%%%%%%
In this section we consider the decision version of $\delta$-dispersion:
\emph{``For a given graph $G=(V,E)$, a positive real $\delta$, and a bound $k$, decide whether
$\disp{\delta}{G}\le k$.''}
Our NP-certificate specifies the following partial information on a $\delta$-dispersed set $S$
in a graph $G=(V,E)$:
\begin{itemize}
\item The certificate specifies the set $W:=V\cap S^*$.
\item For every edge $e\in E$, the certificate specifies the number $n_e$ of facilities
that are located in the interior of $e$.
\end{itemize}
As every edge accommodates at most $1/\delta$ points from $S$, the encoding length of
our certificate is polynomially bounded in the instance size.
For verifying the certificate, we introduce for every vertex $u$ and for every incident 
edge $e=\{u,v\}\in E$ with $n_e>0$ a corresponding real variable $x(u,e)$, which models 
the distance between vertex $u$ and the closest point from $S$ in the interior of edge $e$.
Finally, we introduce the following linear constraints:
\begin{itemize}
\item The non-negativity constraints $x(u,e)\ge0$.
\item For every edge $e=\{u,v\}\in E$, the inequality $x(u,e)+(n_e-1)\delta+x(v,e)\le1$.
\item For all $u,v\in W$ with $u\ne v$, the inequality $d(u,v)\ge\delta$.
\item For all $w\in W$ and $e=\{u,v\}\in E$, the inequality $x(u,e)+d(u,w)\ge\delta$.
\item For all $e=\{u,v\}\in E$ and $e'=\{u',v'\}\in E$, the inequality $x(u,e)+d(u,u')+x(u',e')\ge\delta$.
\end{itemize}
These inequalities enforce that on every edge the variables properly work together,
and that the underlying point set indeed is $\delta$-dispersed.
For verifying the certificate, we simply check in polynomial time whether the resulting linear 
program has a feasible solution, and whether $|W|+\sum_{e\in E}n_e\ge k$ holds.

%%%%%%%%%%%%%%%%%
\begin{theorem}
\label{th:np-containment}
The decision version of $\delta$-dispersion lies in NP, even if the value $\delta$ is 
given as part of the input. \qed
\end{theorem}
%%%%%%%%%%%%%%%%%

%%%%%%%%%%%%%%%%%%%%%%%%%%%%%%%%%%%%%%%%%%%%%%%%%%%%%%%%%%%%%%%%%%%%%%%%%
%%%%%%%%%%%%%%%%%%%%%%%%%%%%%%%%%%%%%%%%%%%%%%%%%%%%%%%%%%%%%%%%%%%%%%%%%
%%%%%%%%%%%%%%%%%%%%%%%%%%%%%%%%%%%%%%%%%%%%%%%%%%%%%%%%%%%%%%%%%%%%%%%%%
\section{The polynomial time result for \texorpdfstring{$\delta=2$}{delta=2}}
\label{sec:delta=2}
%%%%%%%%%%%%%%%%%%%%%%%%%%%%%%%%%%%%%%%%%%%%%%%%%%%%%%%%%%%%%%%%%%%%%%%%%
This section derives a polynomial time algorithm for computing the $2$-dispersion number of a graph.
This algorithm is heavily based on tools from matching theory, as for instance developed in the book 
by Lov\'asz \& Plummer \cite{LovPlu1986}.
As usual, the size of a maximum cardinality matching in graph $G$ is denoted by $\nu(G)$.

%%%%%%%%%%%%%%%%%
\begin{lemma}
\label{le:stefan.0}
Every graph $G=(V,E)$ satisfies~ $\disp{2}{G}\ge\nu(G)$.
\end{lemma}
%%%%%%%%%%%%%%%%%
\proof
The midpoints of the edges in every matching form a $2$-dispersed set.
\qed

\bigskip
A $2$-dispersed set is in \emph{canonical} form, if it entirely consists of vertices and of midpoints of edges.
Recall that by Lemma~\ref{le:halfintegral} every graph $G=(V,E)$ possesses an optimal $2$-dispersed set 
in canonical form.
Throughout this section, we  will consider $2$-dispersed (but not necessarily optimal) sets $S^*$ in
canonical form; we always let $V^*$ denote the set of vertices in $S^*$, and we let $E^*$ denote 
the set of edges whose midpoints are in $S^*$.
Finally, $N^*\subseteq V$ denotes the set of vertices in $V-V^*$ that have a neighbor in $V^*$.
As $S^*$ is $2$-dispersed, the vertex set $V^*$ forms an independent set in $G$, and the edge 
set $E^*$ forms a matching in $G$.
Furthermore, the vertex set $N^*$ separates the vertices in $V^*$ from the edges in $E^*$; 
in particular, no edge in $E^*$ covers any vertex in $N^*$.
We start with two technical lemmas that will be useful in later arguments.

%%%%%%%%%%%%%%%%%
\begin{lemma}
\label{le:stefan.1}
Let $G=(V,E)$ be a graph with a perfect matching, and let 
$S^*$ be some  $2$-dispersed set in canonical form in $G$.
Then $|S^*|\le\nu(G)$. 
\end{lemma}
%%%%%%%%%%%%%%%%%
\proof
Let $M\subseteq E$ denote a perfect matching in $G$, and for every vertex $v\in V$
let $e(v)$ denote its incident edge in matching $M$.
Consider the vertex set $V^*$ and the edge set $E^*$ that correspond to set $S^*$.
Then $E^*$ together with the edges $e(v)$ with $v\in V^*$ forms another matching $M'$ 
of cardinality $|E^*|+|V^*|=|S^*|$ in $G$.
Now $|S^*|=|M'|\le\nu(G)$ yields the desired inequality.
\qed

\bigskip
A graph $G$ is \emph{factor-critical} \cite{LovPlu1986}, if for every vertex $x\in V$ there exists a 
matching that covers all vertices except $x$.
A \emph{near-perfect} matching in a graph covers all vertices in $V$ except one.  
Note that the statement in the following lemma cannot be extended to graphs that consist of a single vertex.
%%%%%%%%%%%%%%%%%
\begin{lemma}
\label{le:stefan.2}
Every $2$-dispersed set $S^*$ in a factor-critical graph $G=(V,E)$ with $|V|\ge3$
satisfies $|S^*|\le\nu(G)$.
\end{lemma}
%%%%%%%%%%%%%%%%%
\proof
Without loss of generality we assume that $S^*$ is in canonical form, and we let
$V^*$ and $E^*$ denote the underlying vertex set and edge set, respectively.
If $V^*$ is empty, we have $|S^*|=|E^*|\le\nu(G)$ since $E^*$ is a matching.
If $V^*$ is non-empty, then also $N^*$ is non-empty (here we use the condition $|V|\ge3$) 
and we pick some vertex $x\in N^*$.
We consider a near-perfect matching $M$ that covers all vertices except $x$, and
we let $e(v)$ denote the edge incident to $v\in V$ in matching $M$.
Then $E^*$ together with the edges $e(v)$ with $v\in V^*$ forms another matching $M'$
of cardinality $|E^*|+|V^*|=|S^*|$ in $G$.
The claim follows from $|S^*|=|M'|\le\nu(G)$.
\qed

%%%%%%%%%%%%%%%%%%%%%%%%%%%%%%%%%%%%%%%%%%%%%%%%%%%%%%%%%%
\begin{figure}[tb]
\centerline{\includegraphics[height=8.0cm]{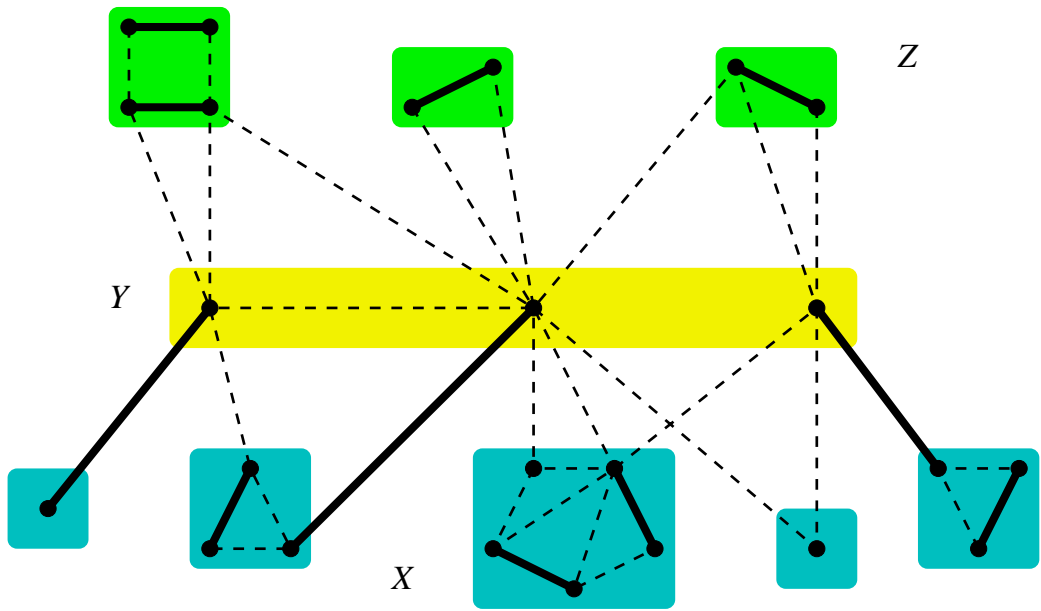}}
\caption{An illustration for the Edmonds-Gallai structure theorem.
A maximum matching is shown with fat edges, and the non-matching edges are dashed.}
\label{fig:EG}
\end{figure}
%%%%%%%%%%%%%%%%%%%%%%%%%%%%%%%%%%%%%%%%%%%%%%%%%%%%%%%%%%

\bigskip
The following theorem goes back to Edmonds \cite{Edmonds1965} and Gallai \cite{Gallai1963,Gallai1964};
see also Lov\'asz \& Plummer \cite{LovPlu1986}.
Figure~\ref{fig:EG} gives an illustration.
%%%%%%%%%%%%%%%%%
\begin{theorem}
\label{th:EG}
(Edmonds-Gallai structure theorem)
Let $G=(V,E)$ be a graph. 
The following decomposition of $V$ into three sets $X,Y,Z$ can be computed in polynomial time.
\begin{eqnarray*}
X &=& \{v\in V\mid \text{ there exists a maximum matching that misses $v$}\} \\[0.5ex]
Y &=& \{v\in V\mid \text{ $v\notin X$ and $v$ is adjacent to some vertex in $X$}\} \\[0.5ex]
Z &=& V-(X\cup Y)
\end{eqnarray*}
The Edmonds-Gallai decomposition has the following properties:
\begin{itemize}
\item 
Set $X$ is the union of the  odd-sized components of $G-Y$; every such odd-sized component is factor-critical.
Set $Z$ is the union of the even-sized components of $G-Y$.
\item Every maximum matching in $G$ induces a perfect matching on every (even-sized) component of $Z$ and
a near-perfect matching on every (odd-sized) component of $X$.
Furthermore, the matching matches the vertices in $Y$ to vertices that belong to $|Y|$ different 
components of $X$.
\qed
\end{itemize}
\end{theorem}
%%%%%%%%%%%%%%%%%

We further subdivide the set $X$ in the Edmonds-Gallai decomposition into two parts:
Set $X_1$ contains the vertices of $X$ that belong to components of size~$1$, and
set $\xodd$ contains the vertices that belong to (odd-sized) components of size at least~$3$.
The \emph{vicinity} $\vic(v)$ of a vertex $v\in V$ consists of vertex $v$ itself and of the
midpoints of all edges incident to $v$.

%%%%%%%%%%%%%%%%%
\begin{lemma}
\label{le:properties}
There exists an optimal $2$-dispersed set $S^*$ in canonical form (with underlying edge set $E^*$) 
that additionally satisfies the following three properties.
\begin{quote}
\begin{itemize}
\item[P1.] In every component of $\xodd$, the set $E^*$ induces a near-perfect matching.
\item[P2.] For every vertex $y\in Y$, the set $\vic(y)\cap S^*$ is either empty or consists
of the midpoint of some edge between $X$ and $Y$.
\item[P3.] In every component of $Z$, the set $E^*$ induces a perfect matching.
\end{itemize}
\end{quote}
\end{lemma}
%%%%%%%%%%%%%%%%%
\proof
We start from an arbitrary optimal $2$-dispersed set $S^*$ (in canonical form, with corresponding 
sets $V^*$ and $E^*$) and transform it in two steps into an optimal $2$-dispersed set of the desired form.

In the first transformation step, we exploit a matching $M$ between sets $Y$ and $X$ that matches 
every vertex $y\in Y$ to some vertex $M(y)$, so that for $y_1\ne y_2$ the vertices $M(y_1)$ and $M(y_2)$ 
belong to different components of $X$; see Theorem~\ref{th:EG}.
A vertex $y\in Y$ is called \emph{blocked}, if it is adjacent to some $x\in X_1\cap S^*$.
As for a blocked vertex the set $\vic(y)\cap S^*$ is already empty (and hence already satisfies property P2), 
we will not touch it at the moment. 
We transform $S^*$ in the following way.
\begin{itemize}
\item
For every non-blocked vertex $y\in Y$, the set $\vic(y)\cap S^*$ contains at most one point.
We remove this point from $S^*$, and we insert instead the midpoint of the edge between $y$ and $M(y)$ into $S^*$.
These operations cannot decrease the size of $S^*$.
\item
Every (odd-sized) component $C$ of $\xodd$ contains at most one point $M(y)$ with $y\in Y$.
We compute a near-perfect matching $M_C$ for $C$ that misses this vertex $M(y)$ (and if no such
vertex is in $C$, matching $M_C$ misses an arbitrary vertex of $C$).
We remove all points in $C$ from $S^*$, and we insert instead the midpoints of the edges in $M_C$.
As by Lemma~\ref{le:stefan.2} we remove at most $\nu(C)$ points and as we insert exactly $\nu(C)$ points,
these operations will not decrease the size of $S^*$.
\end{itemize}
The resulting set $S^*$ is of course again in canonical form, and it is also easy to see that $S^*$ 
is still $2$-dispersed.
Furthermore, $S^*$ now satisfies properties P1 and P2.

In the second transformation step, we note that the current $S^*$ does neither contain vertices from $Y$ nor
midpoints of edges between $Y$ and $Z$.
For every (even-sized) component $C$ of $Z$, we compute a perfect matching $M_C$.
We remove all points in $C$ from $S^*$, and we insert instead the midpoints of the edges in $M_C$.
As by Lemma~\ref{le:stefan.2} we remove at most $\nu(C)$ points and as we insert exactly $\nu(C)$ points,
these operations will not decrease the size of $S^*$.
The resulting set $S^*$ is $2$-dispersed and satisfies properties P1, P2, and P3.
\qed

\bigskip
The optimal $2$-dispersed sets in Lemma~\ref{le:properties} are strongly structured and fairly
easy to understand:
The perfect matchings in set $Z$ contribute exactly $|Z|/2$ points to $S^*$.
Every (odd-sized) component $C$ in $\xodd$ contributes exactly $(|C|-1)/2$ points to $S^*$.
The only remaining open decisions concern the points in $X_1$ and the midpoints of the 
edges $\{y,M(y)\}$ for $y\in Y$.
So let us consider the set $T:=S^*\cap X_1$, and let $\Gamma(T)\subset Y$ denote the vertices
in $Y$ that are adjacent to some vertex in $T$.
Then every vertex $y$ in $Y-\Gamma(T)$ contributes the midpoint of $\{y,M(y)\}$ to $S^*$,
and every vertex $x\in T$ contributes itself to $S^*$.

Hence the remaining optimization problem boils down to finding a subset $T\subseteq X_1$
that maximizes the function value $f(T):=|Y-\Gamma(T)|+|T|$, which is equivalent to
minimizing the function value 
%%%%%%%%%%%%%%%%%
\begin{equation}
\label{eq:submodular}
g(T):=|\Gamma(T)|-|T|.
\end{equation}
%%%%%%%%%%%%%%%%%
The set function $g(T)$ in \eqref{eq:submodular} is a \emph{submodular} function, as it 
satisfies $g(A)+g(B)\ge g(A\cup B)+g(A\cap B)$ for all $A,B\subseteq X_1$; see for instance
\ Gr\"otschel, Lov\'asz \& Schrijver \cite{GroLovSch1988}.
Therefore, the minimum value of $g(T)$ can be determined in polynomial time by the ellipsoid 
method \cite{GroLovSch1988}, or by Cunningham's combinatorial algorithm \cite{Cunningham1985}.

We also describe another way of minimizing the function $g(T)$ in polynomial time, that avoids 
the heavy machinery of submodular optimization and that formulates the problem as a minimum 
$s$-$t$-cut computation in a weighted directed auxiliary graph. 
The auxiliary graph is defined as follows.
\begin{itemize}
\item
Its vertex set contains a source $s$ and a sink $t$, together with all the vertices 
in $X_1$ and all the vertices in $Y$.
\item
For every $x\in X_1$, there is an arc $(s,x)$ of weight $w(s,x)=1$ from the source to $x$. 
For every $y\in Y$,   there is an arc $(y,t)$ of weight $w(y,t)=1$ from $y$ to the sink. 
Whenever the vertices $x\in X_1$ and $y\in Y$ are adjacent in the original graph $G$, 
the auxiliary graph contains the arc $(x,y)$ of weight $w(x,y)=+\infty$.
\end{itemize}
Now let us consider some $s$-$t$-cut of finite weight, which is induced by some vertex 
set $U$ in the auxiliary graph with $s\in U$ and $t\notin U$.
As all arcs from set $X_1$ to set $Y$ have infinite weights, whenever $U$ contains some 
vertex $x\in X_1$ then $U$ must also contain all the neighbors of $x$ in $Y$. 
By setting $T:=X_1\cap U$, we get that the value of the cut equals $|X_1-T|+|\Gamma(T)|$;
hence the minimizer for \eqref{eq:submodular} can be read off the minimizing cut in 
the auxiliary graph.

We finally summarize all our insights and formulate the main result of this section.
%%%%%%%%%%%%%%%%%
\begin{theorem}
\label{th:delta=2}
The $2$-dispersion number of a graph $G$ can be computed in polynomial time.
\qed
\end{theorem}
%%%%%%%%%%%%%%%%%

%%%%%%%%%%%%%%%%%%%%%%%%%%%%%%%%%%%%%%%%%%%%%%%%%%%%%%%%%%%%%%%%%%%%%%%%%
%%%%%%%%%%%%%%%%%%%%%%%%%%%%%%%%%%%%%%%%%%%%%%%%%%%%%%%%%%%%%%%%%%%%%%%%%
%%%%%%%%%%%%%%%%%%%%%%%%%%%%%%%%%%%%%%%%%%%%%%%%%%%%%%%%%%%%%%%%%%%%%%%%%
\section{The polynomially solvable cases}
\label{sec:polynomial}
%%%%%%%%%%%%%%%%%%%%%%%%%%%%%%%%%%%%%%%%%%%%%%%%%%%%%%%%%%%%%%%%%%%%%%%%%
Theorem~\ref{th:delta=2} and Lemma~\ref{le:monotonicity} together imply that for every rational
number $\delta=a/b$ with numerator $a\le2$,  the $\delta$-dispersion number of a graph can be 
computed in polynomial time.
We now present some results that provide additional structural insights into these cases.
The cases where the numerator is $a=1$ are structurally trivial, and the value of the corresponding 
$\delta$-dispersion number can be written down with the sole knowledge of $|V|$ and $|E|$.

%%%%%%%%%%%%%%%%%
\begin{lemma}
\label{le:delta=1}
Let $\delta=1/b$ for some integer $b$, and let $G=(V,E)$ be a connected graph.
\begin{itemize}
\item If $G$ is a tree then $\disp{\delta}{G}=b|E|+1$.
\item If $G$ is not a tree then $\disp{\delta}{G}=b|E|$.
\end{itemize}
\end{lemma}
%%%%%%%%%%%%%%%%%
\proof
If $G$ is a tree, we use a $\delta$-dispersed set $S$ that contains all vertices in $V$
and that for every edge $e=\{u,v\}$ contains all points $p(u,v,i/b)$ with $i=1,\ldots,b-1$.
Clearly $|S|=b|E|+1$.
If $G$ is not a tree, set $S$ contains for every edge $e=\{u,v\}$ all the points
$p(u,v,(2i-1)/(2b))$ with $i=1,\ldots,b$.
Clearly $|S|=b|E|$.

It remains to show that there are no $\delta$-dispersed sets of larger cardinality.
If $G$ is a tree, we root it at an arbitrary vertex so that it becomes an out-tree.
We partition $P(G)$ into $|E|+1$ regions:
One region consists of the root, and all other regions consist of the interior
points on some edge together with the source vertex of that edge.
A $\delta$-dispersed set contains at most $b$ points from every edge-region and at most one
point from the root region.
If $G$ is not a tree, we similarly partition $P(G)$ into $|E|$ regions:
Every region either consists of the interior points of some edge, or of the
interior points of an edge together with one of its incident vertices.
A $\delta$-dispersed set contains at most $b$ points from every such region.
\qed

\bigskip
The following lemma derives an explicit (and very simple) connection between the 
$2$-dispersion number and the $(2/b)$-dispersion number (with odd denominator $b$) of a graph.
The lemma also implies directly that for every odd $b$, the computation of $(2/b)$-dispersion 
numbers is polynomial time equivalent to the computation of $2$-dispersion numbers.
%%%%%%%%%%%%%%%%%
\begin{lemma}
\label{le:numerator=2}
Let $G=(V,E)$ be a graph, let $z\ge1$ be an integer, and let $\delta=2/(2z+1)$. 
Then the dispersion numbers satisfy~ $\disp{\delta}{G}=\disp{2}{G}+z|E|$.
\end{lemma}
%%%%%%%%%%%%%%%%%
\proof
We first show that $\disp{\delta}{G}\ge\disp{2}{G}+z|E|$.
Indeed, let $S_2$ denote an optimal $2$-dispersed set for $G$.
By Lemma~\ref{le:halfintegral} we assume that $S_2$ is in canonical form and hence
entirely consists of vertices and of midpoints of edges.
We partition the edge set $E$ into three parts:
Part $E_1$ contains the edges, for which one end vertex is in $S_2$.
Part $E_{1/2}$ contains the edges whose midpoint lies in $S_2$.
Part $E_0$ contains the remaining edges (which hence are disjoint from $S_2$).
We construct a point set $S_{\delta}\subset P(G)$ as follows:
\begin{itemize}
\item For every edge $\{u,v\}\in E_1$ with $u\in S_2$, we put point $u$ together with
the $z$ points $p(u,v,i\delta)$ with $i=1,\ldots,z$ into $S_{\delta}$.
\item For every edge $\{u,v\}\in E_{1/2}$, we put the $z+1$ points $p(u,v,(4i-3)\delta/4)$
with $i=1,\ldots,z+1$ into $S_{\delta}$.
\item For every $\{u,v\}\in E_0$, we put the $z$ points $p(u,v,(4i-1)\delta/4)$
with $i=1,\ldots,z$ into $S_{\delta}$.
\end{itemize}
It is easily verified that the resulting set $S_{\delta}$ is $\delta$-dispersed and 
contains $|S_2|+z|E|$ points.

Next, we show that $\disp{\delta}{G}\le\disp{2}{G}+z|E|$.
Let $S_{\delta}$ denote an optimal $\delta$-dispersed set for $G$.
By Lemma~\ref{le:halfintegral} we assume that for every point $p(u,v,\lambda)$ in $S_{\delta}$,
the denominator of the rational number $\lambda$ is $2(2z+1)$.
Our first goal is to bring the points in $S_{\delta}$ into a particularly simple constellation.
\begin{itemize}
\item As long as there exist edges $e=\{u,v\}\in E$ with $u,v\in S_{\delta}$, 
we remove all points on $e$ from $S_{\delta}$ and replace them by the $z+1$ points
$p(u,v,(4i-3)\delta/4)$ with $i=1,\ldots,z+1$.
\item Next, for every edge $e=\{u,v\}\in E$ with $u\in S_{\delta}$ and $v\notin S_{\delta}$,
we remove all points on $e$ from $S_{\delta}$ and replace them by the $z+1$ points
$p(u,v,i\delta)$ with $i=1,\ldots,z$.
\item Finally, for every edge $e=\{u,v\}\in E$ with $u,v\notin S_{\delta}$ 
we remove all points on $e$ from $S_{\delta}$ and replace them by the $z$ points 
$p(u,v,(4i-1)\delta/4)$ with $i=1,\ldots,z$.
\end{itemize}
It can be seen that these transformations do not decrease the cardinality of $S_{\delta}$,
and that the resulting set is still $\delta$-dispersed.
Finally, we construct the following set $S_2$ from $S_{\delta}$:
First, $S_2$ contains all points in $V\cap S_{\delta}$,
Secondly, whenever $S_{\delta}$ contains $z+1$ points from the interior of some edge $e\in E$,
then we put the midpoint of $e$ into $S_2$.
It can be shown that the resulting set $S_2$ is $2$-dispersed and has the desired cardinality.
\qed

%%%%%%%%%%%%%%%%%%%%%%%%%%%%%%%%%%%%%%%%%%%%%%%%%%%%%%%%%%%%%%%%%%%%%%%%
%% Bibliography
%%%%%%%%%%%%%%%%%%%%%%%%%%%%%%%%%%%%%%%%%%%%%%%%%%%%%%%%%%%%%%%%%%%%%%%%
\bibliography{stacs}

\end{document}